\documentclass[twocolumn,showpacs,superscriptaddress,preprintnumbers,amsmath,amssymb,prc]{revtex4-2}
\usepackage{multirow}
\usepackage{graphicx}% Include figure files
\usepackage{dcolumn}% Align table columns on decimal point
\usepackage{bm}% bold math
\usepackage{float}
\usepackage{color}
\usepackage{CJK}
\usepackage{ulem}
\usepackage{array} 
\usepackage{makecell}
\usepackage{graphicx}
\usepackage{adjustbox}
\usepackage{appendix}

\usepackage[colorlinks,linkcolor=blue,urlcolor=blue,citecolor=blue]{hyperref}

\usepackage{ragged2e}
\usepackage{tikz,xcolor,hyperref}

\definecolor{lime}{HTML}{A6CE39}
\DeclareRobustCommand{\orcidicon}{
	\begin{tikzpicture}
	\draw[lime, fill=lime] (0,0) 
	circle [radius=0.16] 
	node[white] {{\fontfamily{qag}\selectfont \tiny ID}};
	\draw[white, fill=white] (-0.0625,0.095) 
	circle [radius=0.007];
	\end{tikzpicture}
	\hspace{-2mm}
}
\foreach \x in {A, ..., Z}{%
	\expandafter\xdef\csname orcid\x\endcsname{\noexpand\href{https://orcid.org/\csname orcidauthor\x\endcsname}{\noexpand\orcidicon}}
}
\foreach \x in {A, ..., Z}{%
	\expandafter\xdef\csname orcid\x\endcsname{\noexpand\href{https://orcid.org/\csname orcidauthor\x\endcsname}{\noexpand\orcidicon}}
}

\usepackage{graphicx} % Required for inserting images
\usepackage{amsmath}

\begin{document}
\begin{CJK*}{UTF8}{gbsn}

\title{Bayesian inference of nuclear incompressibility from collective flow in mid-central Au+Au collisions at 400--1500 MeV/nucleon}

\author{J. M. Wang (汪金梅)}
\affiliation{Key Laboratory of Nuclear Physics and Ion-beam Application (MOE), Institute of Modern Physics, Fudan University, Shanghai 200433, China}
\affiliation{Shanghai Research Center for Theoretical Nuclear Physics, NSFC and Fudan University, Shanghai 200438, China}

\author{X. G. Deng (邓先概)\orcidA{}}
\email{xiangai\_deng@fudan.edu.cn}
\affiliation{Key Laboratory of Nuclear Physics and Ion-beam Application (MOE), Institute of Modern Physics, Fudan University, Shanghai 200433, China}
\affiliation{Shanghai Research Center for Theoretical Nuclear Physics, NSFC and Fudan University, Shanghai 200438, China}

\author{W. J.  Xie (谢文杰)}
\affiliation{Department of Physics, Yuncheng University, Yuncheng 044000, China}

\author{B. A. Li (李宝安)\orcidC{}}
\email{Bao-An.Li@Tamuc.edu}
\affiliation{Department of Physics and Astronomy, East Texas A$\&$M University, Texas 75429-3011, USA}

\author{Y. G. Ma (马余刚)\orcidB{}}
\email{mayugang@fudan.edu.cn}
\affiliation{Key Laboratory of Nuclear Physics and Ion-beam Application (MOE), Institute of Modern Physics, Fudan University, Shanghai 200433, China}
\affiliation{Shanghai Research Center for Theoretical Nuclear Physics, NSFC and Fudan University, Shanghai 200438, China}

\date{\today}

\begin{abstract}

The incompressibility $K$ of symmetric nuclear matter (SNM) is determined through a Bayesian analysis of collective flow data from Au + Au collisions at beam energies $E = 400 -1500$  MeV/nucleon. This analysis utilizes a Gaussian process (GP) emulator applied to the isospin-dependent quantum molecular dynamics (IQMD) model for heavy-ion collisions, both with and without incorporating the momentum dependence of the single-nucleon potentials. Specifically, the inferred incompressibility values are $K=188.9^{+2.9}_{-4.5}$ MeV and $256.1^{+8.2}_{-8.7}$ MeV at $E = 400$ MeV/nucleon, respectively, at the 68\% confidence level using rapidity and transverse velocity dependence of proton elliptic flow data, with and without consideration of the momentum dependence. When the transverse momentum dependence of proton-like directed flow data is included, the inferred incompressibility values become $K=222.3^{+9.0}_{-9.9}$ MeV and $K=285.5^{+6.7}_{-7.3}$ MeV, respectively. Furthermore, we found that the value of $K$ derived from observables of proton elliptic flow increases with beam energy. This indicates that the equation of state (EoS) of nuclear matter hardens at higher densities and temperatures in reactions with higher beam energies.

\end{abstract}

\maketitle
\section{Introduction}
\begin{figure*}[htb]
\setlength{\abovecaptionskip}{0pt}
\setlength{\belowcaptionskip}{0pt}
\includegraphics[scale=0.65]{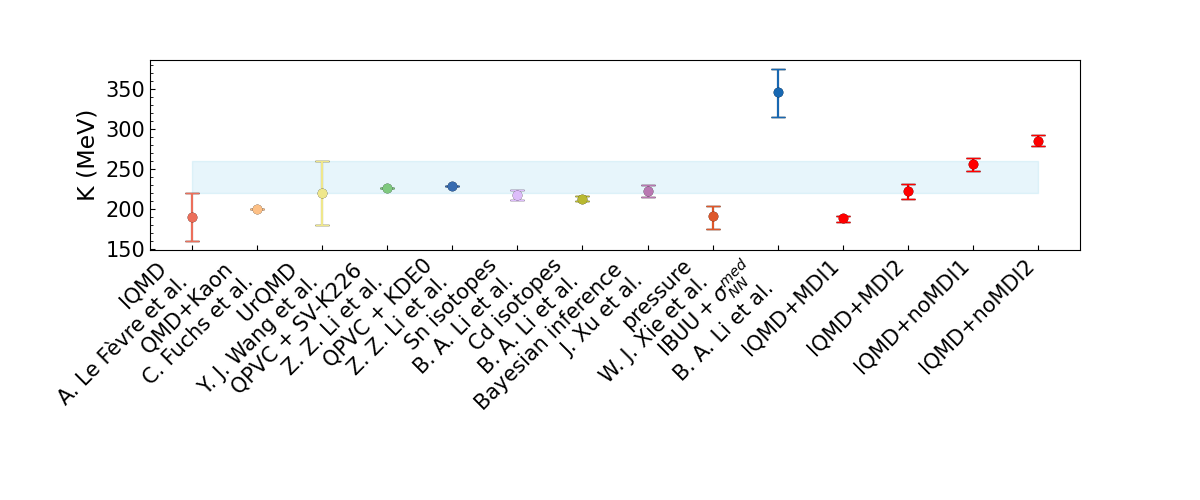}
\vspace{-1.6cm}
\caption{The light blue band covering $K=220-260$ MeV is the fiducial range from studying giant resonances \cite{Garg18,X-Roca-Maza,Blaizot}. The rest are indicated by the legends from left to right from Refs. \cite{A-Le-F-16,C-Fuchs,Y-J-Wang,Z-Z-Li,Z-Z-Li2,Bao-An-Li,J-Xu,WenJie-Xie,B-A-Li}. The red points with error bars are the results of $K$ {at $E = 400$ MeV/nucleon} from this work.}
\label{fig:fig1}
\end{figure*}

The equation of state (EoS) of cold symmetric nuclear matter (SNM) is a fundamental relationship between the energy per nucleon $E/A$ and nucleon density $\rho$. At the saturation density $\rho_0=0.16 $ $\mathrm{fm^{-3}}$ of SNM, where $E/A = -16.0$ MeV and pressure vanishes, the stiffness of SNM EoS is measured by its incompressibility $K =9\rho ^{2}\frac{\partial ^{2}E/A }{\partial \rho ^{2} }\mid_{\rho _{0} }$. Pinning down precisely the value of $K$ has been a longstanding and shared goal of both nuclear physics and astrophysics for the last few decades because of its strong impact on many aspects of nuclear structure and reactions as well as properties of neutron stars, mechanisms of supernovae explosions and emissions of gravitational waves from mergers of neutron stars, see, e.g., Refs.~\cite{Blaizot, P-Moller, P-Senger, G-F-Burgio21, J-B-Wei, G-f-Burgio20,S-Balberg, M-Oertel, D-Chatterjee, D-Gerstung,Wang20,Ma23,J-Y-Xu, H-Shen, J-M-Lattimer,Xu2,Zhang,ZHWucpc1,wjXiecpc2,AMZhaocpc3}. In fact, thanks to the hard work of many people, especially in the last 40 years, much progress has been made in constraining the value of $K$. In particular, since the pioneering work of Blaizot who determined $K$ = (210 $\pm$ 30) MeV from analyzing the giant monopole resonance (GMR) energies in $^{\rm 40}$Ca, $^{\rm 90}$Zr and $^{\rm 208}$Pb \cite{Blaizot}, extensive theoretical studies and systematic measurements of GMR energies \cite{Blaizot,You,Garg18,X-Roca-Maza,Jorge10,Stone,Colo14} have led to the community consensus that the $K$ has values in the range of 220 MeV to 260 MeV \cite{Garg18,X-Roca-Maza,Colo14,shlomo06} or around $235\pm 30$ MeV \cite{Khan1,MM1} as indicated by the light blue band in Fig. \ref{fig:fig1}. 

Another well-known tool for studying the EoS of nuclear matter is heavy-ion collisions. In particular, various components of nuclear collective flow (e.g., directed and elliptic flow) as well as yields and spectra of multiple particles (e.g., pions, kaons, photons and dileptons) are useful probes of the EoS albeit with different sensitivities \cite{Reisdorf,Heinz,MaYG}. Information about the EoS is normally extracted by comparing transport model simulations with experimental data in the forward-modeling approach, while in more recent years the data-driven Bayesian inference has become more revealing with quantified uncertainties. In either approach, there are still many interesting issues in narrowing down the remaining uncertainties of various features and parameters of the nuclear EoS using heavy-ion reactions, e.g., Refs. \cite{Li2008PR,TMEP,Sorensen} for reviews. Heavy-ion collisions are also widely used to study the equation of state of nuclear matter at higher baryon density in recent works~\cite{S-Huth,D-Oliinychenko,C-Y-Tsang,M-O-Kuttan,Agnieszka-Sorensen2021}.

The purpose of this work is  twofold. Firstly, uncertainties remain in the values of $K$ derived both from  analyses of GMR energies and certain transport model studies of heavy-ion reactions. For example, as shown by the first point in Fig.~\ref{fig:fig1}, analyses of the elliptic flow in Au+Au reactions at beam energies from 0.4 to 1.5 GeV/nucleon from the FOPI collaboration within the Isospin-dependent Quantum Molecular Dynamics (IQMD) model extracted a value of $K=190^{+30}_{-30}$ MeV~\cite{A-Le-F-16}. Furthermore, an analysis of the ratio of kaons in Au+Au over C+C systems within hadronic transport simulations favors a soft EoS of $K=200$ MeV than the hard EoS of $K=380$ MeV~\cite{C-Fuchs}. Both studies have considered the momentum dependence of isoscalar single-nucleon potentials. For more other studies about the nuclear incompressibility $K$, see Fig.~\ref{fig:fig1}. It should be noted that the incompressibility $K$ in Refs.~\cite{A-Le-F-16,C-Fuchs,Y-J-Wang,WenJie-Xie,B-A-Li} is obtained from transport models of heavy-ion reactions in which $\rho > \rho _{0} $. And in Refs.~\cite{Z-Z-Li,Z-Z-Li2,Bao-An-Li,J-Xu}, it is derived from studies of the GMR which is $\rho \approx  \rho _{0}$. The uncertainty of the approximately 40 MeV error band of $K$ from studying GMR and the $K$ from analyzing heavy-ion reactions are still too large for many investigations in both nuclear physics and astrophysics. For example, it has been shown recently that the variation of $K$ between 220 and 260 MeV leads to significant changes in the crust-core transition density and pressure in neutron stars \cite{LiM,Richter}. They subsequently affect the radii, tidal polarizabilities, and crustal moments of inertia of canonical neutron stars, hindering the investigations of the very mysterious high-density cores of neutron stars as their properties are often strongly correlated with the value of $K$. Thus, a re-extraction of the $K$ with a quantified uncertainty from heavy-ion collisions in a Bayesian approach will be invaluable. Considering that the momentum dependence of single-nucleon potentials is very important for accurately simulating heavy-ion reactions, we have inferred the value of $K$ with or without considering the momentum dependence of single-nucleon potentials.

Secondly, we investigate the inference of $K$ from different observables and different beam energies used. By studying the combined data of rapidity and transverse velocity dependence of proton elliptic flow in mid-central Au+Au reactions at 400 MeV/nucleon, we found an incompressibility $K=188.9^{+2.9}_{-4.5}$ or $256.1^{+8.2}_{-8.7}$ MeV at 68\% confidence level with or without considering the momentum dependence of single-nucleon potentials, respectively. As shown in Fig.~\ref{fig:fig1}, these results are all consistent with the earlier findings of Refs. \cite{A-Le-F-16,C-Fuchs,Y-J-Wang}. On the other hand, by incorporating the transverse momentum dependence of directed flow of proton-like, the inferred $K$ becomes $222.3^{+9.0}_{-9.9}$ MeV or $285.5^{+6.7}_{-7.3}$ MeV at 68\% confidence level with or without considering the momentum dependence of single-nucleon potentials, respectively. Obviously, these results are significantly different from those inferred earlier, indicating that the difference of the extracted EoS is due to the different density regions probed by different observable. While we have not studied extensively all observables in a broad beam energy range yet, our findings reported here are useful for unraveling the physics underlying the incompressibility of nuclear matter and the constraining powers of different observables useful for eventually pinning down the value of $K$. Furthermore, through the analysis of proton elliptic flow in mid-central Au+Au reactions at beam energies of $400-1500$ MeV/nucleon, with consideration of the momentum dependence of single-nucleon potentials, we found that the value of $K$ increases with beam energy, reflecting the hardening of nuclear matter EoS at higher densities and temperatures in reactions with higher beam energies.

The rest of this article is organized as follows: in Sec. \ref{sec:IQMD model}, we outline the key ingredients of the IQMD model most relevant for this study. In Sec.~\ref{sec:Bayesian approach}, we recall the key aspects of the Bayesian approach used in this work. In Sec. \ref{sec:Gaussian process emulator}, we present the Gaussian Process (GP) emulator of IQMD, and demonstrate the accuracy of its training and testing. The posterior probability distribution functions (PDFs) of the incompressibility $K$ as well as the corresponding nuclear interaction parameters from our Bayesian inference in different situations are presented and discussed in Sec. \ref{sec:results}. In Sec. \ref{sec:Conclusions}, we summarize the results from our work.

\section{A brief summary of the IQMD model for heavy-ion reactions}\label{sec:IQMD model}

For ease of understanding and completeness, we recall here some key physical ingredients of the IQMD model that are most relevant to this work. This model has been widely and successfully applied to the study of heavy ion collisions at low to intermediate beam energies, see for example Refs. \cite{J-Aichelin1,J-Aichelin,C-Hartnack,Zhang1,Zhang2,Wei2024NST,Wang2023NST,Xiao2023NST}. The nuclear effective interactions used in our QMD model includes not only a Skyrme term, but also the Yukawa, isospin asymmetric and Coulomb terms. Optionally, a momentum dependent interaction (MDI) term can be switched on or off. More specifically, the EoS, pressure and incompressibility $K$ for cold SNM can be written as \cite{J-Aichelin1,L-P-Csernai,G-Giuliani,Y-K-Vermani,J-Aichelin,C-Hartnack}:

\begin{equation}
\begin{split}
E/A& = \frac{\alpha }{2}\frac{\rho }{\rho _{0} }+\frac{\beta }{\gamma +1}{\left ( \frac{\rho }{\rho _{0} }  \right ) ^{ \gamma }}+\frac{3}{10m}\left ( \frac{3\pi ^{2}\hbar ^{3} \rho}{2}  \right ) ^{2/3}\\
&+\frac{1}{2}t_{4}\frac{\rho }{\rho _{0} }\int f\left ( \vec{p}  \right )\ln^{2}\left [ 1+t_{5}\left ( \vec{p}-\left \langle {\vec{p}}'  \right \rangle \right ) ^{2}   \right ]d^{3}p ,     
\end{split}
\label{EOS1}
\end{equation}

\begin{equation}
\begin{split}
P&=\rho ^{2}\frac{\partial E/A}{\partial \rho }  =\frac{\alpha }{2 } \frac{\rho ^{2}  }{\rho _{0} }+\frac{\beta \gamma \rho }{\gamma +1}\left ( \frac{\rho }{\rho _{0} }  \right ) ^{\gamma }  +\frac{1}{5m}\left ( \frac{3}{2}\pi ^{2}\hbar ^{3}    \right ) ^{\frac{2}{3} } \rho ^{\frac{5}{3} }\\&+ \frac{t_{4} }{2}\frac{\rho ^{2}  }{\rho _{0} } \ln^{2}\left ( 1+t_{5}P_{F}^{2}   \right ) ,  
\end{split}
\label{P1}
\end{equation}

\begin{equation}
\begin{split}
K& =9\rho ^{2}\frac{\partial ^{2}E/A }{\partial \rho ^{2} }\mid_{\rho _{0} }=-\frac{3}{5m}\left ( \frac{3\pi ^{2}\hbar^{3}\rho _{0}   }{2}  \right ) ^{2/3}+\frac{9\beta \gamma \left ( \gamma -1 \right ) }{\gamma +1}\\
&+\ln{\left ( 1+t_{{5} } P_{F} ^{2}  \right ) }\frac{6t_{4}t_{5} P_{F}^{2} }{1+t_{{5} } P_{F}^{2} }, 
\end{split}
\label{K1}
\end{equation}
where $\alpha$, $\beta$, $\gamma$, $t_{4} $ and $t_{5} $ are parameters, $P_{F}=\left(3\pi^{2} \hbar^{3} \rho/2\right)^{1/3}$ is the nucleon Fermi momentum at density $\rho$, and $f\left(\vec{p}\right ) =\frac{3}{4\pi P_{F}^{3} }\Theta \left ( p _{F}-p   \right )$, $\Theta $ is the step function. The values of $t_{4}=1.57$ MeV and $t_{5}=5\times 10^{-4} $ $\mathrm{c^{2}/MeV^{2}}$ are determined by the experimental nucleon optical potential and the isoscalar nucleon effective mass at $\rho_0$ \cite{J-Aichelin1}. The EoS and its characteristics depend only on $\alpha$, $\beta$, and $\gamma$ parameters. In this work, a default free-space nucleon-nucleon cross section, $\sigma _{NN}^{free}$ obtained from experimental data，is used for the simulations \cite{F-Daffin1996,A-B-Larionov2001}.

\section{Bayesian approach for inferring EoS parameters}\label{sec:Bayesian approach}
Bayesian inference is a probabilistic approach employed in Machine Learning (ML) that has been widely utilized in many fields in recent years. Compared to the traditional $\chi^{2}$ fitting in the forward-modeling approach, it has several advantages in determining model parameters with quantified uncertainties, see, e.g. Refs.~\cite{zhangzhen2021,Pratt,Scott,Bass2,Bass3,W-B-He1,W-b-He2,K-Zhou,B-A-Li,M-O-Kuttan,LGP,P-Morfouace,N-K-Patra,J-Xu,Cox,Xie1,Zhou,Alh,Xie2,Alq}. For completeness, we recall that the Bayes' theorem states that
\begin{equation}
\begin{split}
P\left ( \theta \mid D \right )=\frac{P\left ( D \mid \theta \right )P\left ( \theta  \right ) }{P\left ( D \right ) },
\end{split}
\end{equation}
where $P\left(\theta \mid D\right)$ denotes the posterior probability density function (PDF) of model parameter set $\theta$ given the dataset $D$, while $P\left(D \mid \theta\right)$ is the likelihood for the model with the parameter set $\theta$ to reproduce the dataset $D$. The $P\left(\theta\right)$ is the prior PDF of the parameter set $\theta$, and $P\left(D\right)$ serves as the normalization constant. In this study, the $\theta$ consists of incompressibility $K$, $\alpha$, $\beta$ and $\gamma$ constrained by the three SNM saturation conditions at $\rho_0$ mentioned earlier. Thus, only one parameter is free, and we choose it to be $K$. In the Markov Chain Monte Carlo (MCMC) process of our Bayesian analysis, the trial $K$ value is randomly generated uniformly in the prior range of $K = 170 \sim 420$ MeV. Once the incompressibility $K$ is selected, the corresponding values of $\alpha$, $\beta$, and $\gamma$ are then also obtained from the saturation conditions. Additionally, two constraints of $-1000<\alpha <0$ MeV and $\beta >0$ MeV are applied to keep nuclear matter stable and to remain casual at all densities, which in turn raises the lower bound of the prior distribution of $K$ to above 170 MeV. This can be seen in the prior distribution of $K$ shown in Sec.~\ref{sec:results}. Thus, there is a one-to-one correspondence between the incompressibility $K$ and $\alpha$, $\beta$, $\gamma$ parameters considering the saturation conditions of SNM at $\rho_0$ through Eqs.~(\ref{EOS1})--(\ref{K1}). 
By inputting a given set of parameters $\alpha$, $\beta$, and $\gamma$ into the IQMD model, theoretical predictions for the reaction observables corresponding to the selected EoS parameter set can be obtained. These predictions will then be used in evaluating the likelihood function as we shall discuss next.

Taking the logarithm of the posterior distribution of parameters enables a transformation of the Bayesian formula \cite{M-O-Kuttan}:
\begin{equation}
\begin{split}
\mathrm{ln}\left ( P\left ( \theta \mid D \right ) \right ) \propto \mathrm{ln}\left ( P\left ( D \mid \theta \right ) \right ) +\mathrm{ln}\left ( P\left ( \theta  \right ) \right ).
\end{split}
\end{equation}
The logarithm of the likelihood function is evaluated by using
\begin{equation}
\begin{split}
\mathrm{ln}\left ( P\left ( D \mid \theta \right ) \right ) = -\frac{1}{2}\sum_{i}\left [ \frac{\left (y _{i}^{\theta }-y_{i,exp}   \right ) ^{2} }{\sigma _{i}^{2} } +\mathrm{ln}\left ( 2\pi \sigma _{i}^{2}  \right )  \right ].  
\end{split}
\end{equation}
In the above, $\sigma _{i} = \sqrt{\sigma_{exp}^{2} +\sigma_{mod}^{2}}$ where $\sigma_{exp}$ and $\sigma_{mod}$ (evaluated from the emulator) represent the experimental and model errors, respectively. The $y _{i}^{\theta }$ and $y_{i,exp}$ are the model prediction for the observable $y_i$ and its experimental value, respectively.

\section{Training and testing a Gaussian process (GP) emulator of the IQMD simulator}\label{sec:Gaussian process emulator}
Since it is impractical to invoke the computationally expensive IQMD simulator in Bayesian analyses, its emulator must be used instead. Here we use the popular and well-tested GP emulator. Below we provide some details on the training and testing of the emulator.

In this study, we use four observables in Au + Au collisions at $E = 400$ MeV/nucleon {and two observables at $E = 600-1500$ MeV/nucleon}, namely, the rapidity dependence of $-v_{2}\left(y_{0}\right)$ of proton elliptic flow for $u_{t0}>0.4$, the transverse velocity dependence of $-v_{2}\left(u_{t0}\right)$ of proton elliptic flow for $\left | y_{0} \right | <0.4$~\cite{FOPI_Collaboration}, and the transverse momentum dependence of $v_{1} \left ( p_{t}^{\left ( 0 \right ) }  \right ) $ of proton-like (all charged particles with $Z=1$ and $Z=2$ are weighted by the charge Z) directed flow in two rapidity ranges of $y_{0}=0.5-0.7$ and $y_{0}=0.7-0.9$~\cite{FOPI_Collaboration1}. For the proton elliptic flow, the centrality $0.25 < b_{0} < 0.45$ is considered, where $b_{0}$ denotes the scaled impact parameter defined as $b_{0} = b/b_{\text{max}}$, with $b_{\text{max}} = 1.15(A_{P}^{1/3} + A_{T}^{1/3})$ fm. The range of impact parameter $1.9-6.1$ fm is considered for proton-like directed flow. Here, $v_{2} = \langle\cos 2\phi\rangle = \langle (p_{x}^{2} - p_{y}^{2}) / (p_{x}^{2} + p_{y}^{2}) \rangle$. The reduced rapidity $y_{0}$ is defined as $y/y_{\text{pro}}$, where $y = \frac{1}{2}\ln\left(\frac{E+p_{z}}{E-p_{z}}\right)$, and `pro' denotes the incident projectile in the center of mass frame. The transverse (spatial) component $u_{t}$ of the 4-velocity $u$ is given by $u_{t} = \beta_{t} \gamma _{c} $. Here, the 3-vector $\vec{\beta}$ represents the velocity in units of the speed of light, and $\gamma _{c}  = 1/\sqrt{1-\beta^{2}}$. The unitless transverse velocity $u_{t0} = u_{t} / u_{\text{pro}}$ where $u_{\text{pro}} = \beta_{\text{pro}} \gamma_{\text{pro}}$ is used. For the directed flow, $v_{1} = \langle\cos \phi\rangle =\langle p_{x} / \sqrt{p_{x}^{2} + p_{y}^{2}} \rangle$, we study its dependence on the normalized center-of-mass (c.m.) transverse momentum (per nucleon) is defined as $p_{t}^{\left ( 0 \right ) } =\left ( p_{t}/A  \right ) /\left ( p_{P}^{\mathrm{c.m.}}/A_{P}   \right ) $. To understand the role of observables on the constrained result of $K$, we do comparative studies using two datasets: one group utilizes only the rapidity dependence of proton elliptic flow $-v_{2}\left(y_{0}\right)$ and its transverse velocity dependence $-v_{2}\left(u_{t0}\right)$, while the other incorporates the transverse momentum dependence of $v_{1} \left ( p_{t}^{\left ( 0 \right ) }  \right )$ of proton-like directed flow in two rapidity ranges. Additionally, we utilize only the observables of proton elliptic flow $-v_{2}\left(y_{0}\right)$ and $-v_{2}\left(u_{t0}\right)$ to investigate the beam energy dependence of the constrained results for the incompressibility $K$.

\begin{figure}[htb]
\setlength{\abovecaptionskip}{0pt}
\setlength{\belowcaptionskip}{8pt}
\includegraphics[scale=0.95]{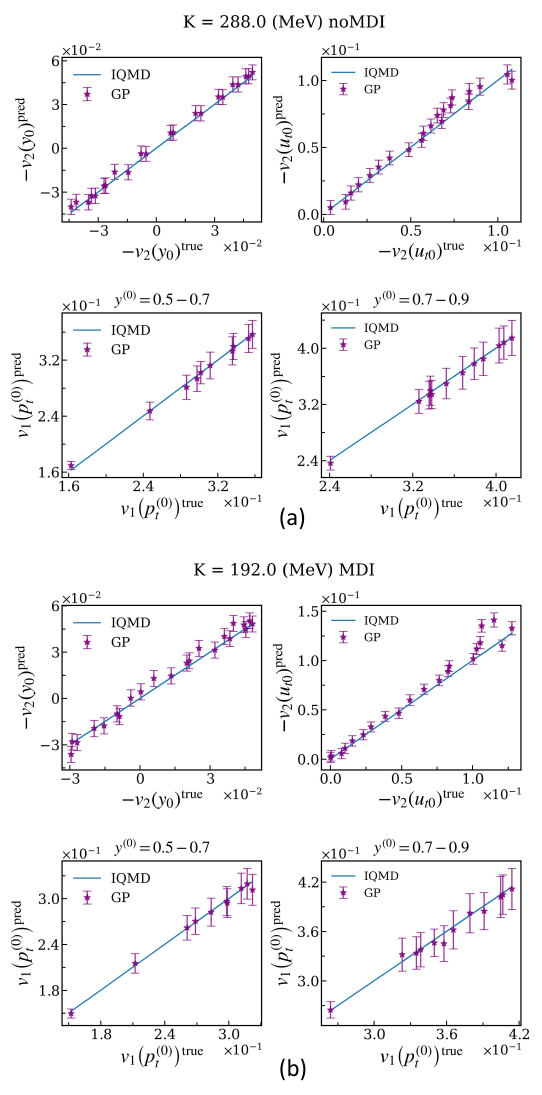}
\caption{A comparison between the predictions for $-v_{2}\left ( y_{0}  \right )  $, $-v_{2}\left ( u_{t0}  \right )  $ and $v_{1} \left ( p_{t}^{\left ( 0 \right ) }  \right )$ by the IQMD simulator and the GP emulator for two randomly selected sets of incompressibility $K$ in the testing dataset. Up: without considering the MDI, down: considering the MDI.}
\label{fig:fig2}
\end{figure}

\begin{figure*}[htb]
\setlength{\abovecaptionskip}{0pt}
\setlength{\belowcaptionskip}{8pt}
\includegraphics[scale=0.6]{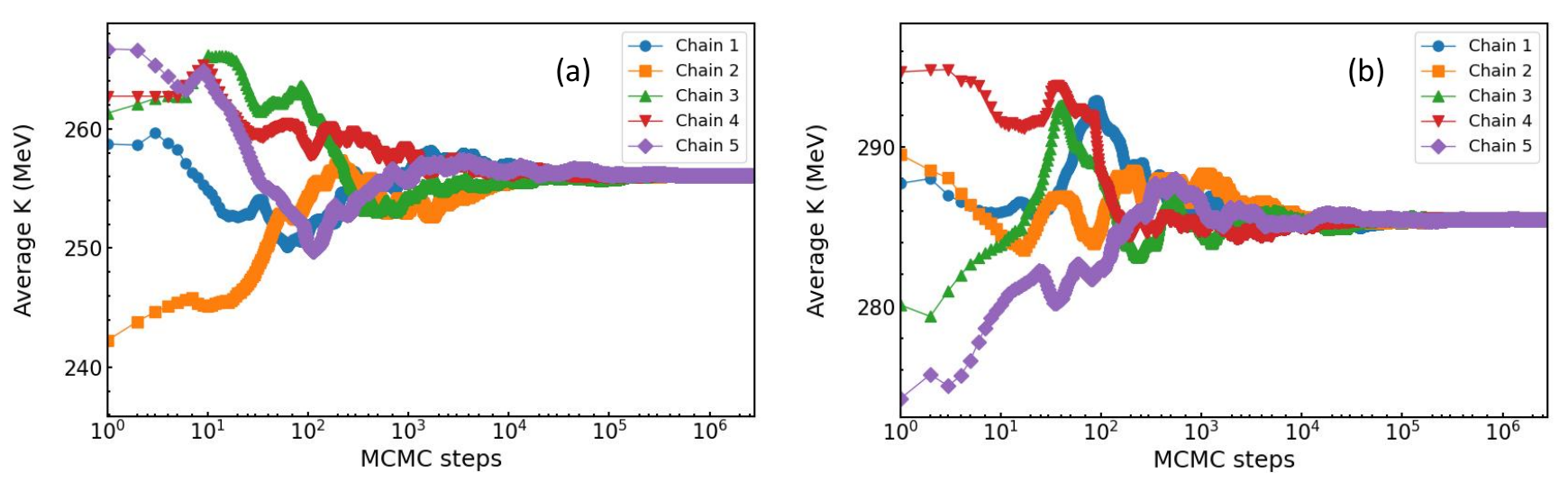}
\caption{Without considering the MDI: running averages of the $K$ as functions of the MCMC steps after 200,000 initial steps in 5 independent chains. Left: observables $-v_{2}\left ( y_{0}  \right )  $ and $-v_{2}\left ( u_{t0}  \right )  $, right: observables $-v_{2}\left ( y_{0}  \right )$, $-v_{2}\left ( u_{t0}  \right )  $ and $v_{1} \left ( p_{t}^{\left ( 0 \right ) }  \right )$.}
\label{fig:fig3}
\end{figure*}

\begin{figure*}[htb]
\setlength{\abovecaptionskip}{0pt}
\setlength{\belowcaptionskip}{8pt}
\includegraphics[scale=0.68]{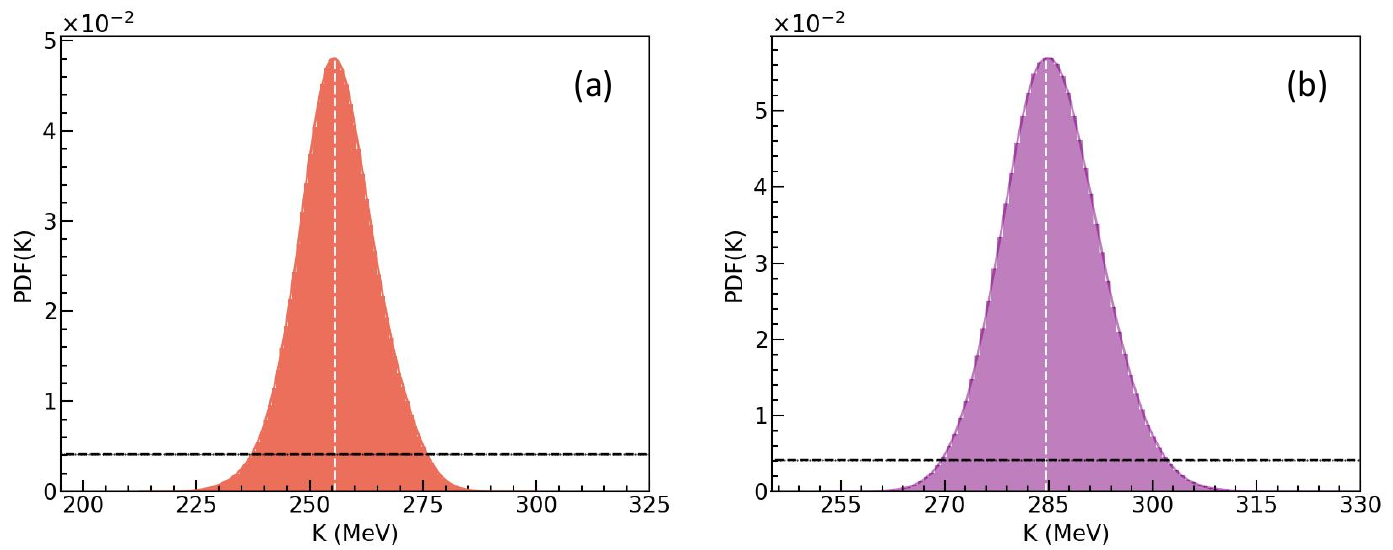}
\caption{Without considering the MDI: the posterior PDFs of $K$. Left: using the observables $-v_{2}\left ( y_{0}  \right )$ and $-v_{2}\left ( u_{t0}  \right )  $, right: observables $-v_{2}\left ( y_{0}  \right )$, $-v_{2}\left ( u_{t0}  \right )  $ and $v_{1} \left ( p_{t}^{\left ( 0 \right ) }  \right )$.}
\label{fig:fig4}
\end{figure*}

\begin{figure*}[htb]
\setlength{\abovecaptionskip}{0pt}
\setlength{\belowcaptionskip}{8pt}
\includegraphics[scale=0.64]{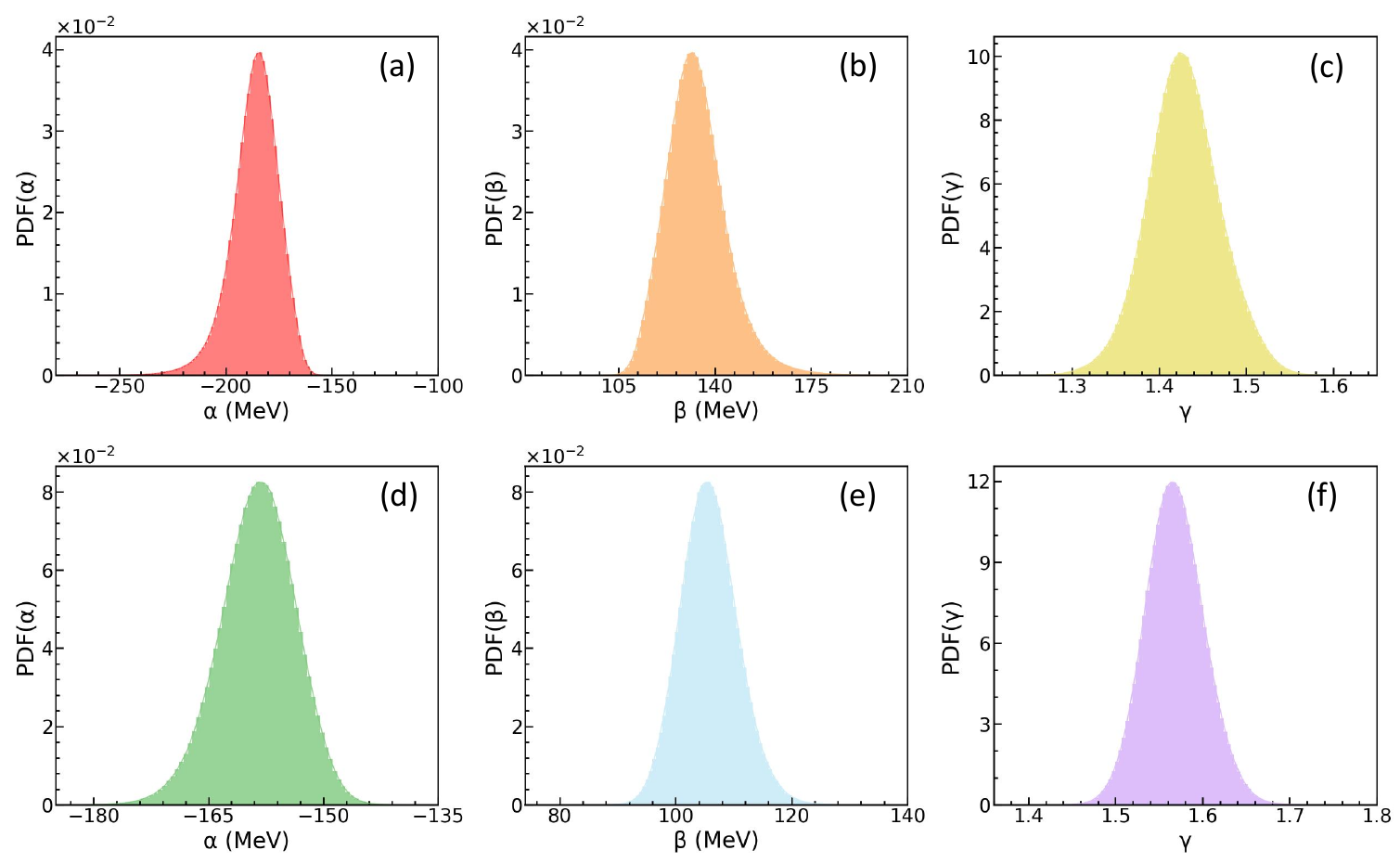}
\caption{Without considering the MDI: the posterior PDFs of $\alpha$, $\beta$ and $\gamma$. Up: observables $-v_{2}\left ( y_{0}  \right )  $ and $-v_{2}\left ( u_{t0}  \right )  $, down: observables $-v_{2}\left ( y_{0}  \right )$, $-v_{2}\left ( u_{t0}  \right )  $ and $v_{1} \left ( p_{t}^{\left ( 0 \right ) }  \right )$.}
\label{fig:fig5}
\end{figure*}

\begin{figure*}[htb]
\setlength{\abovecaptionskip}{0pt}
\setlength{\belowcaptionskip}{8pt}
\includegraphics[scale=0.68]{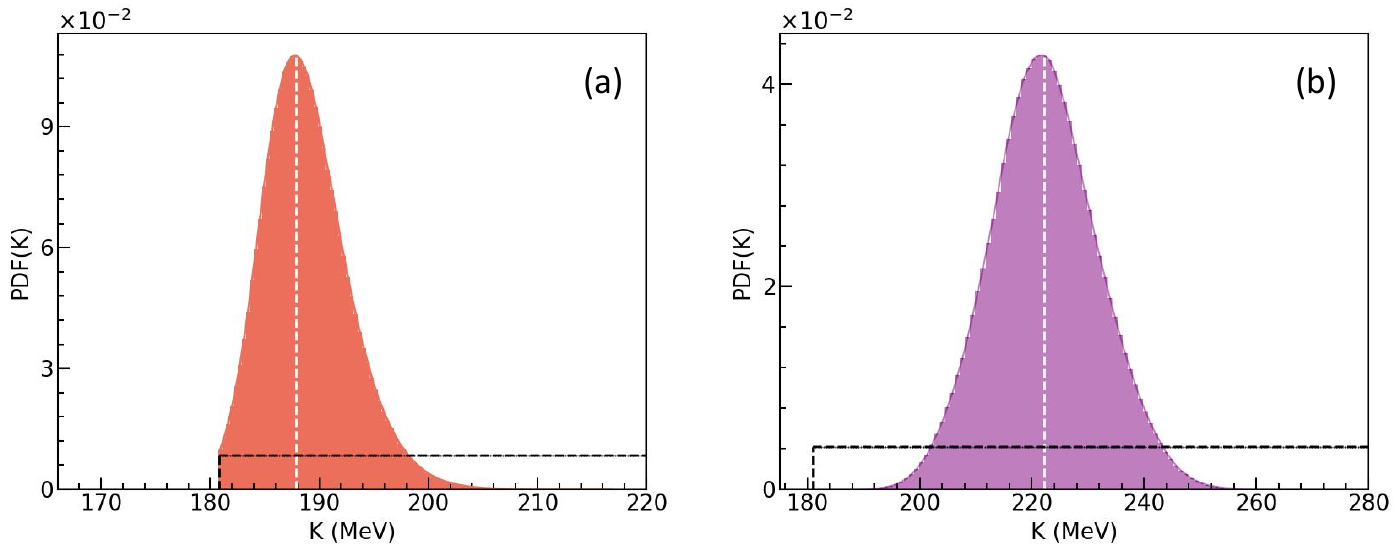}
\caption{Considering the MDI: the posterior PDFs of $K$. Left: observables $-v_{2}\left ( y_{0}  \right )  $ and $-v_{2}\left ( u_{t0}  \right )  $, right: observables $-v_{2}\left ( y_{0}  \right )$, $-v_{2}\left ( u_{t0}  \right )  $ and $v_{1} \left ( p_{t}^{\left ( 0 \right ) }  \right )$.}
\label{fig:fig6}
\end{figure*}

\begin{figure*}[htb]
\setlength{\abovecaptionskip}{0pt}
\setlength{\belowcaptionskip}{8pt}
\includegraphics[scale=0.64]{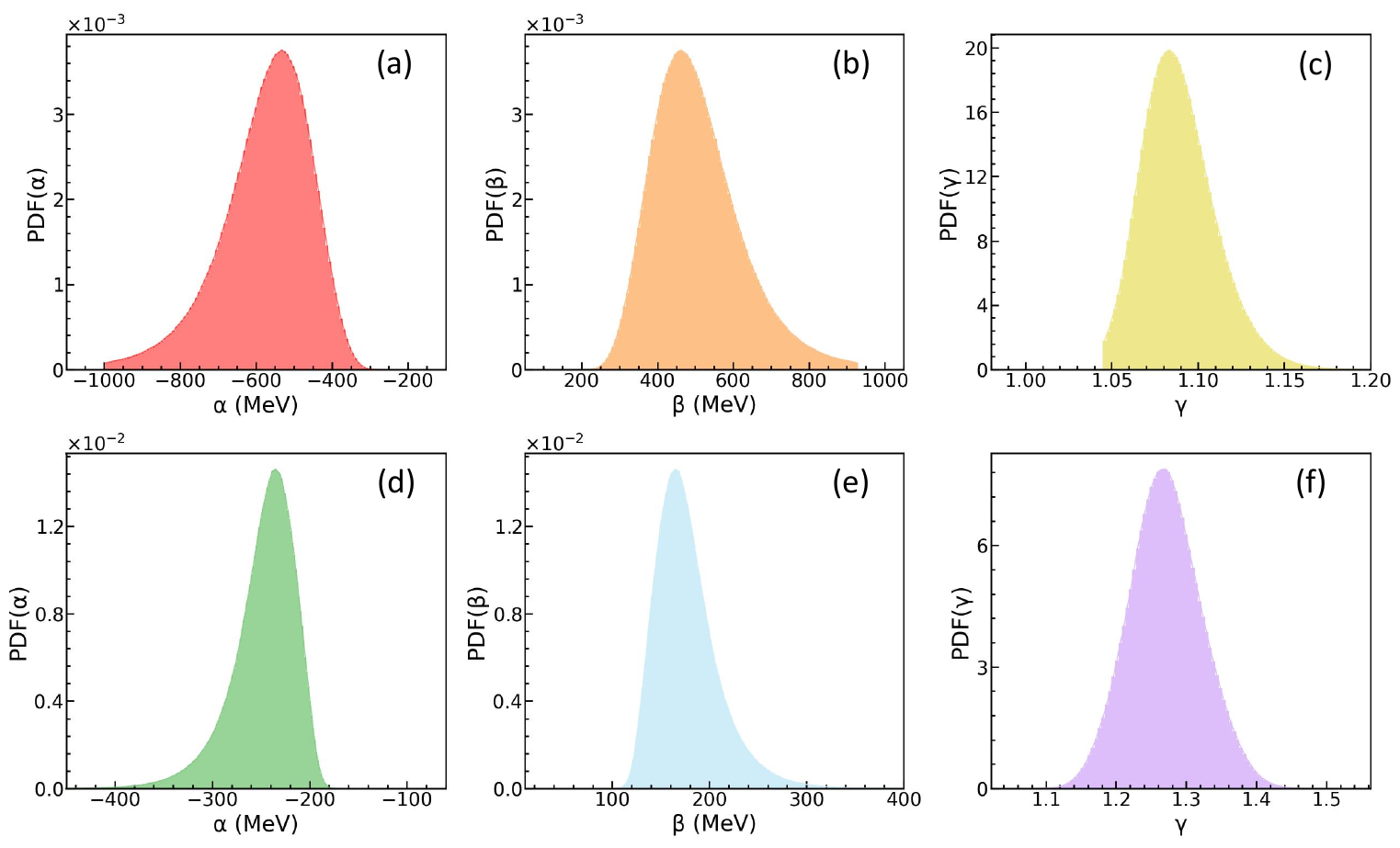}
\caption{Considering the MDI: the posterior PDFs of $\alpha$, $\beta$ and $\gamma$. Up: observables $-v_{2}\left ( y_{0}  \right )  $ and $-v_{2}\left ( u_{t0}  \right )  $, down: observables $-v_{2}\left ( y_{0}  \right )$, $-v_{2}\left ( u_{t0}  \right )  $ and $v_{1} \left ( p_{t}^{\left ( 0 \right ) }  \right )$.}
\label{fig:fig7}
\end{figure*}

\begin{figure*}[htb]
\setlength{\abovecaptionskip}{0pt}
\setlength{\belowcaptionskip}{8pt}
\includegraphics[scale=0.64]{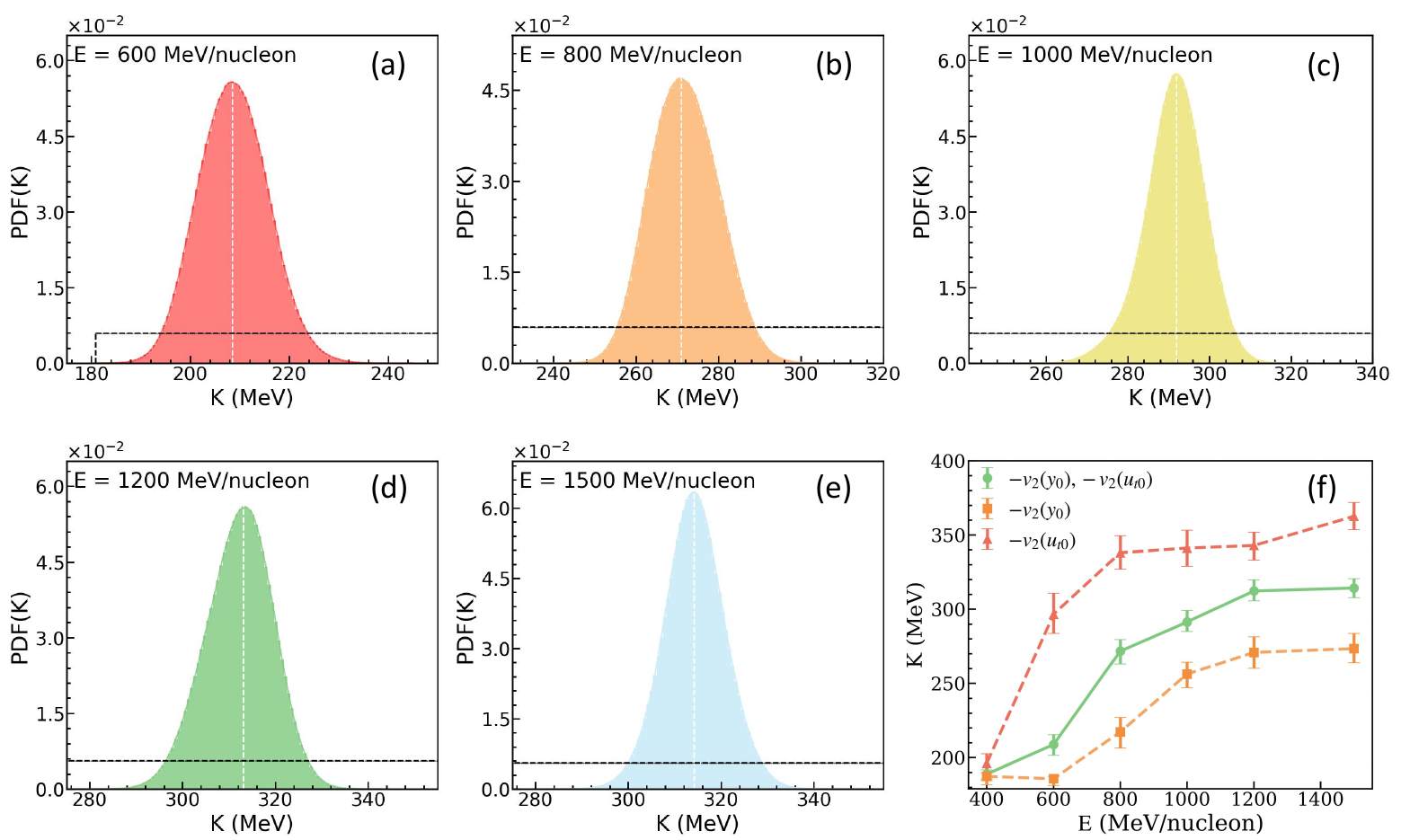}
\caption{Considering the MDI: the posterior PDFs of $K$ at $E=600-1500$ MeV/nucleon and its dependence on beam energy. Observables: $-v_{2}\left ( y_{0}  \right )  $ and $-v_{2}\left ( u_{t0}  \right )  $.}
\label{fig:fig8}
\end{figure*}

%%%%%%%%%%%%%%%%%%%%%%%%%%%%%

The initial value of incompressibility $K$ is selected by using the Latin hypercube sampling (LHS) \cite{M-O-Kuttan,B-A-Li} to cover its entire prior range uniformly and efficiently, enabling the emulator to efficiently learn and reliably predict reaction observables. In generating the training and testing sets with the IQMD simulator, 220 different $K$ values form the training set, while an additional 50 different $K$ values are used for testing purposes. For each incompressibility $K$, the IQMD model produces 100,000 events to calculate the respective observables $-v_{2}\left(y_{0}\right)$, $-v_{2}\left(u_{t0}\right)$ and $v_{1} \left ( p_{t}^{\left ( 0 \right ) }  \right )$, with each event being associated with a randomly selected impact parameter in the range mentioned earlier.

To facilitate effective learning and reliable predictions of observables by the emulator, the commonly employed Radial Basis Function (RBF) is utilized \cite{gaussianprocess}. To assess the emulator's performance on the testing set, shown in Fig.~\ref{fig:fig2} is a visual comparison between model calculations (x axis) by the IQMD simulator and the GP emulator predictions with error bars (y axis), and the line of perfect match between the GP predictions and the IQMD simulations is blue. As evident in Fig.~\ref{fig:fig2}, the error bars given by GP emulator are near or over the blue line, so the GP emulator can basically accurately predict the observables. Clearly, for all the observables in the cases considered, the emulator's performance is satisfactory. 

%%%%%%%%%%%%%%%%%%%%%%%%%%%%%%

\section{Results and Discussions of Bayesian Analyses}\label{sec:results}
The Affine Invariant MCMC Ensemble sampler algorithm is employed to sample the posterior distribution of incompressibility $K$. The cumulative mean (running average) plot from the MCMC sampling is typically employed to assess the convergence status of the sampling process. In this work, each of the five MCMC chains runs independently 3 million steps and discarding the first 1.2 million burn-in steps.

As mentioned earlier, we compare two calculations with or without considering the momentum-dependent interaction (MDI) part of single-nucleon potentials {at $E = 400$ MeV/nucleon}. In both cases, we found that 1 million burn-in steps are sufficient. As an example, shown in Fig.~\ref{fig:fig3}(a) and \ref{fig:fig3}(b) are the cumulative mean diagrams of our MCMC sampling without considering the MDI. It is seen that regardless of the initial states, all chains surely converge to the same equilibrium state after about 1 million steps. Moreover, this conclusion is independent of the observables we used. All results presented in the following are obtained by using 3 million steps, discarding the first 1.2 million burn-in steps.

We first examine the posterior PDFs without considering the MDI. The results obtained from the convergent MCMC chains are shown in Fig.~\ref{fig:fig4}(a) (using only $-v_{2}\left (y_{0}  \right )$ and $-v_{2}\left ( u_{t0}  \right )$) and \ref{fig:fig4}(b) (using $-v_{2}\left ( y_{0}  \right )$, $-v_{2}\left ( u_{t0}  \right )$ and $v_{1} \left ( p_{t}^{\left ( 0 \right )}  \right )$). We note that the white dashed line corresponds to the most probable value (MPV) of the posterior distribution. The black dashed line represents the prior distribution of $K$. In both cases, the posterior PDF is significantly narrower than the prior, demonstrating that the chosen observables provide strong constraints on $K$.

It is important to note that here we adopt the highest posterior density (HPD) method to calculate the 68\% confidence interval~\cite{N-Turkkan} since the posterior distribution of $K$ sometimes deviates from a Gaussian distribution. More specifically, we find from the PDF shown in Fig.~\ref{fig:fig4}(a) that the mean value of $K$ is $K=256.1^{+8.2}_{-8.7}$ MeV at 68\% confidence interval when only using the observables $-v_{2}\left(y_{0}\right)$ and $-v_{2}\left ( u_{t0}  \right )$. On the other hand, from Fig.~\ref{fig:fig4}(b), we find $K=285.5^{+6.7}_{-7.3}$ MeV when $-v_{2}\left(y_{0}\right)$, $-v_{2}\left(u_{t0}\right)$ and $v_{1} \left ( p_{t}^{\left ( 0 \right ) }  \right )$ are used. In this case, the mean or MPV of $K$ is appreciably larger than that when only the elliptic flow data of proton $-v_{2}\left(y_{0}\right)$ and $-v_{2}\left(u_{t0}\right)$ are used. This demonstrates that different combinations of observables, including the proton elliptic flow and the directed flow of proton-like, impose distinct constraints on the incompressibility $K$, as will be discussed in the following.

As previously discussed, given an incompressibility $K$, the corresponding parameters $\alpha$, $\beta$, and $\gamma$ can be derived using the equations for $E/A$, $P$, and $K$ at the saturation density $\rho_0$. Utilizing the convergent MCMC chains, we can readily obtain the posterior PDFs of $\alpha$, $\beta$, and $\gamma$. As shown in Fig.\ref{fig:fig5}(a)--(f), these posterior PDFs are significantly narrower than their prior distributions, indicating tighter constraints. Such constrained parameters are valuable for testing fundamental interactions, such as the Skyrme nuclear effective interactions.

We now turn to the results obtained when incorporating the MDI. We maintain the same prior distributions for incompressibility $K$ and apply identical constraining conditions as in the previous case. The PDFs for $K$ are depicted in Fig.~\ref{fig:fig6}(a) and \ref{fig:fig6}(b). The corresponding results for $\alpha$, $\beta$ and $\gamma$ are illustrated in Fig.~\ref{fig:fig7}(a)--\ref{fig:fig7}(c) and \ref{fig:fig7}(d)--\ref{fig:fig7}(f), respectively. Notably, the mean values of $K$ are now $188.9^{+2.9}_{-4.5}$ MeV at 68\% confidence level when only the observables $-v_{2}\left(y_{0}\right)$ and $-v_{2}\left(u_{t0}\right)$ are used, and $222.3^{+9.0}_{-9.9}$ MeV when the observables $v_{1} \left ( p_{t}^{\left ( 0 \right ) }  \right )$ are incorporated. These results reveal a clear dependence of the inferred $K$ on the choice of observables, similar to the earlier findings. Furthermore, the values of $K$ obtained when including the MDI are significantly smaller than those inferred without considering the MDI. To reproduce the same observables, an equivalent gradient of mean-field potential is required, which can originate from either the density-dependent or momentum-dependent components of the mean-field potential. The MDI potential is repulsive, which provides a larger gradient of mean-field potential for particles without making the density-dependent part of EoS stiffer~\cite{BALI}. However, the absence of a momentum-dependent potential in the simulation requires compensation from a stiffer density-dependent EoS to reproduce the same observables. Thus, in the case of including the momentum-dependent potential in simulating heavy-ion reactions, regardless of the observables used only the $-v_{2}\left(y_{0}\right)$ and $-v_{2}\left(u_{t0}\right)$ or including the $v_{1} \left ( p_{t}^{\left ( 0 \right ) }  \right )$, the required $K$ values are small than without considering the MDI. The results here are consistent with the earlier findings that the MDI reduces the necessary $K$ to reproduce flow data, see, e.g. Refs.~\cite{A-Le-F-16,C-Fuchs,C-Gale,P-Danielewicz,P-Danielewicz1}. However, we acknowledge that the extracted values of $K$ are still subject to uncertainties arising from model dependence. As discussed extensively in the Transport Model Evaluation Project (TMEP) review ~\cite{TMEP} and references therein, there are systematic differences in various transport models.

\begin{table*}[ht]
  \centering 
  \setlength{\tabcolsep}{10pt} 
  \caption{The mean (at 68\% confidence level) and the most probable value of the incompressibility $K$ (MeV) inferred with and without considering the momentum-dependent interaction.} 
  \scalebox{1}{
  \label{tab:tab1}
\begin{tabular}{ccccc}
    \hline 
     &\multicolumn{2}{c}{noMDI ($E=400$ MeV/nucleon)}&\multicolumn{2}{c}{MDI ($E=400$ MeV/nucleon)}\\
    \hline
    \multirow{2}{*}{Observables:} & \multirow{2}{*}{$-v_{2}\left ( y_{0}  \right ),-v_{2}\left ( u_{t0}  \right )$} & {$-v_{2}\left ( y_{0}  \right ),-v_{2}\left ( u_{t0}  \right ),$}  & \multirow{2}{*}{$-v_{2}\left ( y_{0}  \right ),-v_{2}\left ( u_{t0}  \right )$} & {$-v_{2}\left ( y_{0}  \right ),-v_{2}\left ( u_{t0}  \right ),$}\\
    &  & $v_{1} \left ( p_{t}^{\left ( 0 \right ) }  \right )$ & &$v_{1} \left ( p_{t}^{\left ( 0 \right ) }  \right )$ \\
    $K$ Mean  &~~$256.1^{+8.2}_{-8.7}$& $285.5^{+6.7}_{-7.3}$& $188.9^{+2.9}_{-4.5}$& $222.3^{+9.0}_{-9.9}$\\
    $K$ MPV &255.5 & 284.6&187.9&222.3\\
    \hline
    &\multicolumn{3}{c}{MDI (observables:$-v_{2}\left ( y_{0}  \right ),-v_{2}\left ( u_{t0}  \right )$)}\\
    \hline
    600 MeV/nucleon &800 MeV/nucleon & 1000 MeV/nucleon&1200 MeV/nucleon&1500 MeV/nucleon\\
    $208.8^{+6.7}_{-7.3}$ & $271.8^{+7.8}_{-8.7}$&$291.4^{+7.8}_{-6.3}$ & $312.3^{+7.6}_{-6.6}$& $314.3^{+6.3}_{-6.5}$\\
    208.5 &270.9 &291.9 & 313.1&314.1\\
    \hline
    
  \end{tabular}
  }
\end{table*}

The results presented earlier were obtained using proton elliptic flow and proton-like directed flow in mid-central Au+Au collisions at 400 MeV/nucleon. For a systematic study, we now infer the incompressibility solely based on the rapidity and transverse velocity dependence of proton elliptic flow in mid-central Au+Au collisions at beam energies from 600 to 1500 MeV/nucleon, with the momentum-dependent interaction taken into account. Maintaining the same prior distribution of incompressibility $K$ and the same constraining conditions as before. The beam energy dependence of the PDFs of $K$ are illustrated in Fig.~\ref{fig:fig8}(a)--\ref{fig:fig8}(e), respectively. To be more quantitative, the mean values of $K$ as functions of beam energy is showing in Fig.~\ref{fig:fig8} (f) and the values of $K$ are given in Table \ref{tab:tab1}. The incompressibility $K$ increases with beam energy, regardless of whether the observable $-v_{2}\left(y_{0}\right)$ or $-v_{2}\left(u_{t0}\right)$ is used. From Eq.(3), it is clear that the incompressibility \( K \) is defined at the saturation density \( \rho_0 \), and in principle, \( K \) as a property of nuclear matter at \( \rho_0 \) should not depend on the beam energy or the specific observable. However, in transport model simulations where the EoS is characterized by a single parameter \( K \), the observable-sensitive density regime probed by heavy-ion collisions increases with beam energy. Therefore, the \( K \) values extracted from data at different beam energies effectively reflect the stiffness of the EoS at different densities, rather than the incompressibility at \( \rho_0 \). The observed increasing trend of extracted \( K \) values with beam energy should thus be interpreted as an indication that the EoS becomes stiffer at higher densities and temperatures in this energy region from 400 to 1500 MeV/nucleon. And it is consistent with the conclusion from the SMASH model in Ref.~\cite{LATarasovicova2024}. The general trend also aligns with Li's recent work~\cite{Li-arxiv2025}, despite differences in the models and observables used. While it is opposite to the conclusion which gives a soft EoS with elliptic flow in the same energy region in Le F\'{e}vre work~\cite{A-Le-F-16,ALeF2018}. The observable used in this work differs from that in Refs.~\cite{A-Le-F-16,ALeF2018}. Moreover, it should be noted that the description of the $u_{t0}$ cut condition adopted from Ref.~\cite{FOPI_Collaboration} in Ref.~\cite{ALeF2018} is inaccurate, as the descriptions of the $u_{t0}$ cut in Refs.~\cite{FOPI_Collaboration} and~\cite{FOPI3} are inconsistent. This inconsistency may affect the rigor of the conclusions drawn in Ref.~\cite{ALeF2018}. Notably, the values of $K$ derived from the observables $-v_{2}\left(y_{0}\right)$ and $-v_{2}\left(u_{t0}\right)$ exhibit discrepancies, which arise from the differences in the rapidity and transverse velocity dependence of proton elliptic flow. The reason here is similar to the inferred values of $K$ at $E=400$ MeV/nucleon: when different observables probe different density regions, the extracted equation of state will accordingly differ. Specifically, elliptic flow may reflect the behavior of particles at low densities, while directed flow might be more sensitive to high-density behavior. As a result, the EoS extracted from proton elliptic flow tends to be softer, whereas including directed flow leads to a stiffer EoS, as illustrated in Fig.~\ref{fig:fig4} and Fig.~\ref{fig:fig6}.

\section{Summary}\label{sec:Conclusions}

In summary, within a Bayesian statistical framework using a Gaussian process emulator for the IQMD simulator, we have inferred the incompressibility $K$ of the SNM from the proton elliptic flow and proton-like directed flow data in mid-central Au+Au reactions at 400{-1500} MeV/nucleon from the FOPI collaboration. Compared to previous work, which was mostly based on transport model simulations in the forward modelling approach, Bayesian analyses allow us to infer the underlying transport model parameters with quantified uncertainties. In particular, we infer an incompressibility of $K=188.9^{+2.9}_{-4.5}$ MeV or $K=256.1^{+8.2}_{-8.7}$ MeV with 68\% confidence from the combined proton elliptic flow data {at $E=400$ MeV/nucleon} in analyses considering the momentum dependence of the single-nucleon potentials or not. It is consistent with results from previous analyses of the same data using forward modelling, but with significantly reduced uncertainties. However, we acknowledge that the derived $K$ values are still subject to model dependence, an issue that warrants further investigation.

Our study also showed that the incompressibility $K$ derived from heavy-ion reactions is related to the observable-sensitive density regime. Specifically, the derived values of $K$ are $222.3^{+9.0}_{-9.9}$ MeV or $285.5^{+6.7}_{-7.3}$ MeV {at $E=400$ MeV/nucleon}, depending on whether both proton elliptic flow and proton-like directed flow are considered, with or without accounting for the momentum dependence of the single-nucleon potentials. This finding may suggest that the elliptic flow reflects the behavior of particles at low densities, while the directed flow is more sensitive to high-density behavior. Thus, the EoS extracted from proton elliptic flow tends to be softer, whereas including directed flow leads to a stiffer EoS. Furthermore, the inferred values of $K$ derived from the observables of proton elliptic flow in Au+Au collisions at beam energies ranging from 400 to 1500 MeV/nucleon reveal a general trend of progressive hardening of nuclear matter EoS at higher densities and temperatures.

\begin{acknowledgments}
We would like to thank Prof. Kai Zhou for useful communications. Wang, Deng and Ma were supported in part by the National Natural Science Foundation of China under contract Nos. $12147101$, $12347149$, $11890714$, and $11925502$, and the Strategic Priority Research Program of CAS under Grant No. XDB34000000. Xie was supported in part by the Shanxi Provincial
Foundation for Returned Overseas Scholars under Grant No. 20220037, the Natural Science Foundation of Shanxi Province under Grant No. 20210302123085, and the discipline construction project of Yuncheng university. B.A. Li acknowledges support by the U.S.
Department of Energy, Office of Science, under Award No. DE-SC0013702, and the CUSTIPEN (China-U.S. Theory Institute for Physics with Exotic Nuclei) under U.S. Department of Energy Award No. DE-SC0009971.
\end{acknowledgments}

\end{CJK*}
\end{document}